\documentclass[aps,prx,twocolumn,secnumarabic]{revtex4-1} 

\usepackage[utf8]{inputenc}
\usepackage{longtable}
\usepackage{morefloats}
\usepackage[dvips]{graphicx,color}
\usepackage[dvipsnames]{xcolor}
\usepackage{epsfig,graphicx,amsfonts,amsbsy}
\usepackage{graphicx,amsfonts,amsbsy}
\usepackage{amsmath,amsfonts,amsthm,amssymb}
\usepackage{url}
\usepackage{verbatim}
\usepackage{array}
\usepackage{multirow}
\usepackage{tabularx}
\usepackage{multirow}
\usepackage{braket} 
\usepackage{dsfont}
\usepackage{bm}
\usepackage{natbib}
\usepackage{multibib}
\usepackage[colorlinks=true, pdfstartview=FitV, linkcolor=blue,
citecolor=blue, urlcolor=blue]{hyperref}
\usepackage[capitalise]{cleveref}

\newcommand{\ex}{\mathcal{J}}
\newcommand{\zfsd}{\mathcal{D}}
\newcommand{\zfse}{\mathcal{E}}

\begin{document}

\title{Anomalous zero-field splitting  for hole spin qubits in Si and Ge quantum dots}
\author{Bence Het\'enyi}
\email{bence.hetenyi@unibas.ch}
\author{Stefano Bosco}
\author{Daniel Loss}
\affiliation{Department of Physics, University of Basel, Klingelbergstrasse 82, CH-4056 Basel, Switzerland} 
\date{\today}

\begin{abstract}
An anomalous energy splitting of spin triplet states at zero magnetic field has recently been measured in germanium quantum dots.
This zero-field splitting could crucially alter the coupling between tunnel-coupled quantum dots, the basic building blocks of state-of-the-art spin-based quantum processors, with profound implications for semiconducting quantum computers.
We develop an analytical model linking the zero-field splitting to spin-orbit interactions that are cubic in momentum. Such interactions naturally emerge in hole nanostructures, where they can also be tuned by external electric fields,  and we find them to be particularly large in silicon and germanium, resulting in a significant zero-field splitting in the $\mu$eV range.
We confirm our analytical theory by numerical simulations of different quantum dots, also including other possible sources of zero-field splitting. Our findings are applicable to a broad range of current architectures encoding spin qubits and provide a deeper understanding of these materials, paving the way towards the next generation of  semiconducting quantum processors.
\end{abstract}

\maketitle

\paragraph*{Introduction.} 

The compatibility of localized spins in semiconducting quantum dots (QDs)~\cite{loss1998quantum} with the well-developed CMOS technology is pushing these architectures to the front of the race towards the implementation of scalable quantum computers~\cite{eriksson2004spinbased,burkard2021semiconductor,bellentani2021toward,kloeffel2013prospects,scappucci2020germanium}. Spin qubits based on hole states in silicon (Si) and germanium (Ge), in particular, are gaining increasing attention in the community~\cite{kloeffel2013prospects,scappucci2020germanium} because of their large spin-orbit interaction (SOI)~\cite{kloeffel2011strong,terrazos2018theory,froning2021strong,liu2022gatetunable}, enabling fast and power-efficient all-electric gates~\cite{froning2021ultrafast, camenzind2021spin,wang2022ultrafast} and strong transversal and longitudinal coupling to microwave resonators~\cite{kloeffel2013cqed,mutter2021natural,michal2022tunable,bosco2022fully,bosco2022hole}. Also, significant steps forward in material engineering~\cite{hendrickx2018gatecontrolled,scappucci2021crystalline} as well as fast spin read-out and qubit initialization protocols~\cite{vukusic2018single,watzinger2018germanium,urdampilleta2019gatebased,hendrickx2022singleshot} facilitated the implementation of high-fidelity two-qubit gates~\cite{greilich2011optical,hendrickx2020fast} and of a four-qubit quantum processor with controllable qubit-qubit couplings~\cite{hendrickx2020four}.

In contrast to electrons, the properties of hole QDs depend on the mixing of two bands,  the heavy-hole (HH) and light-hole (LH) bands, resulting in several unique features that are beneficial for quantum computing applications~\cite{kloeffel2018direct,winkler2008cubic,moriya2014cubic,hetenyi2020exchange,li2020holephonon,bosco2020hole,secchi2020inter,bosco2021fully}. In addition to the large and externally controllable SOI~\cite{kloeffel2011strong,kloeffel2018direct,bosco2020hole}, that can be conveniently engineered to be linear or cubic in momentum~\cite{bulaev2007electric,winkler2008cubic,bosco2021squeezed,terrazos2018theory,wang2021optimal,piot2022single},  hole spin qubits also feature highly anisotropic and electrically tunable $g$-factors~\cite{maier2013tunable,crippa2018electrical,venitucci2018electrical,studenikin2019electrically,qvist2022anisotropic},  hyperfine interactions~\cite{bosco2021fully}, and anisotropies of exchange interaction at finite magnetic fields~\cite{hetenyi2020exchange}.
Because HHs and LHs are strongly mixed in quasi one-dimensional (1D) systems, these effects are significantly enhanced in long QDs.

Recent experiments in Ge QDs with even hole occupation  have also detected a large anomalous lifting of the threefold degeneracy of triplet states at zero magnetic field~\cite{katsaros2020zero}, yielding another striking difference between electrons and holes. A similar zero-field splitting (ZFS)  has been reported in other quantum systems e.g., divacancies in silicon carbide~\cite{davidsson2018first}, nitrogen-vacancies in diamond~\cite{lenef1996electronic,maze2011properties}, and carbon nanotubes~\cite{szakacs2010zero}, where it is associated to the anisotropy of the two-particle exchange interaction. In this letter, we discuss the microscopic origin of this anisotropy in hole QDs and we propose a general theory modelling the ZFS in a wide range of devices. Our theory helps to develop a fundamental understanding of ZFS, essential to account for its effect in quantum computing applications. 
For example, the exchange anisotropy could enable the encoding of hole singlet-triplet qubits~\cite{jirovec2021singlet,jirovec2022dynamics,mutter2021allelectrical,fernandez2022quantum} at zero magnetic filed, and when combined with a Zeeman field, it can lift the Pauli spin-blockade, with critical implications in read-out protocols~\cite{ono2002current}. Furthermore,  ZFS can introduce systematic errors in two-qubit gates based on isotropic interactions between tunnel-coupled QDs~\cite{loss1998quantum,burkard1999coupled,hetenyi2020exchange}.

We associate the large ZFS emerging in hole QDs to a SOI cubic in momentum. 
The SOI is a natural candidate to explain exchange anisotropies, however, its dominant contribution --linear in momentum-- can be gauged away in quasi 1D systems~\cite{braun1996,levitov2003dynamical,stepanenko2012singlet} and cannot lift the triplet degeneracy without magnetic fields. While in electronic systems only the linear SOI is sizeable, in hole nanostructures the large mixing of HHs and LHs induces a large cubic SOI~\cite{winkler2008cubic,moriya2014cubic} yielding a significant ZFS in Si and Ge QDs.
Strikingly, this ZFS is tunable by external electric fields and can be engineered by the QD design. 

We develop a theory for the cubic-SOI induced ZFS that relies exclusively on single-particle properties of the QD and the Bohr radius, providing an accurate estimate of the ZFS in a wide range of common architectures. In realistic systems, this ZFS is in the $\mu$eV range, orders of magnitude larger than alternative mechanisms. For example, we find that ZFS of a few neV can also be induced by short-range corrections of the Coulomb interaction arising from the $p$-type orbital wavefunctions of the valence band~\cite{secchi2020inter,satoko1990calculated}. In addition, our theory relates the axis of the exchange anisotropy to the direction of the SOI, and corroborates the observed response of the QDs to small magnetic fields~\cite{katsaros2020zero}. Importantly, because in long QDs comprising two holes the Coulomb repulsion of the two particles forms a double QD~\cite{taut1993two,abadillo2021wigner, gao2020site,ercan2021strong}, our theory describes the exchange anisotropy also in tunnel-coupled QDs, the prototypical building blocks of current spin-based quantum processors~\cite{burkard1999coupled,hetenyi2020exchange,secchi2021interacting}, and thus our findings have profound implications in the growing research field of quantum computing with holes.

\begin{figure}
\centering
\includegraphics[width=0.8\columnwidth]{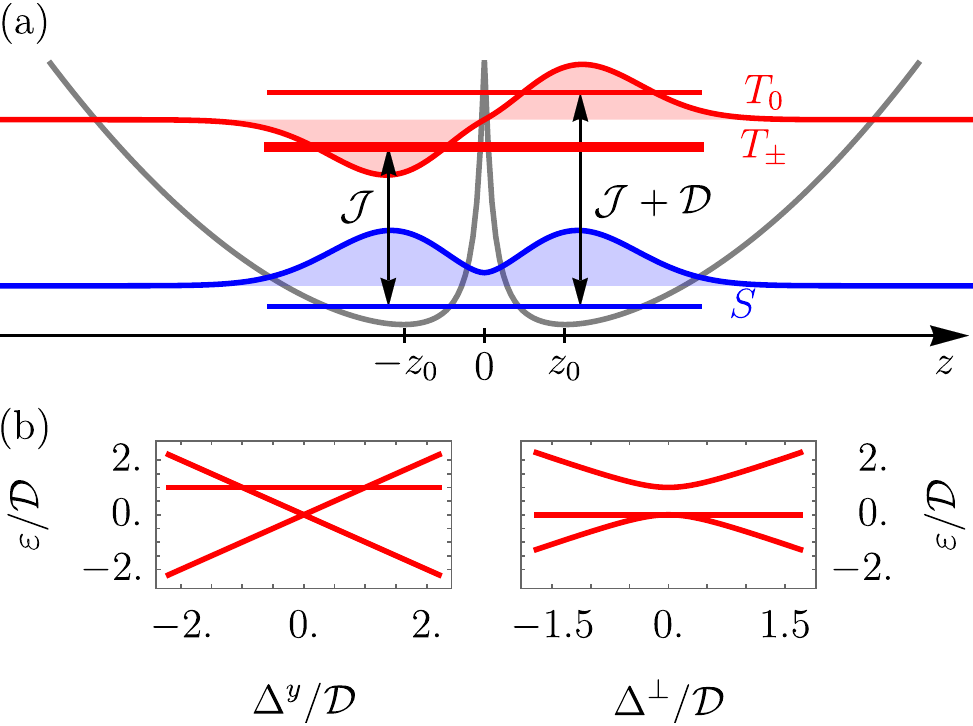}
\caption{Exchange interaction in long quantum dots.
(a) The effective 1D potential $V_c(z_1-z_2)$ is shown in gray (without units) as a function of relative coordinate $z = z_1-z_2$, where $\pm z_0$ are the minima of the potential. The energy levels corresponding to the lowest singlet and triplet states, and the corresponding orbital wavefunctions are overlayed with blue and red, respectively. Vertical arrows show the definition of the exchange splitting and ZFS, $\ex$ and $\zfsd$, respectively. Note that the energy scale of the singlet-triplet energy levels is only schematic, not matched with that of the effective potential. (b) Splitting $\epsilon$ of the three triplet states when the Zeeman field is aligned with the SOI ($\Delta^y$, left panel), and when it is  perpendicular to it ($\Delta^{\perp}$, right panel). }
\label{fig:Effpot_splittings}
\end{figure} 

\begin{figure*}
\centering
\includegraphics[width=0.95\textwidth]{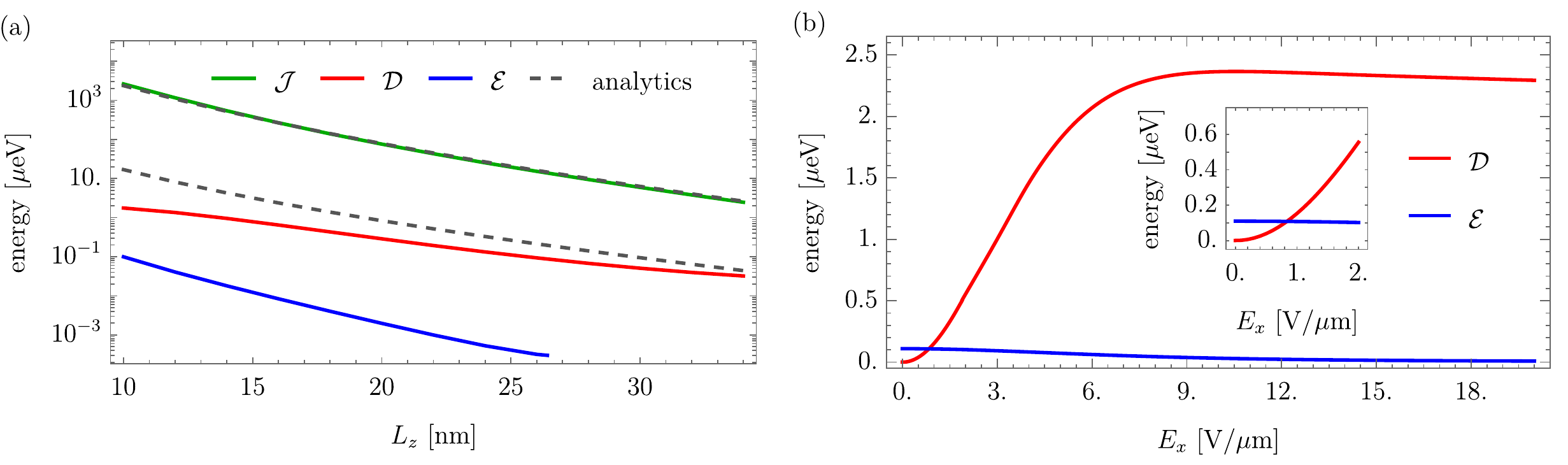}
\caption{Anisotropic exchange interactions in Ge. (a) Exchange splitting $\mathcal J$ and ZFSs $\mathcal D$ and $\mathcal E$ in a Ge square NW with  side length $L = 10\,$nm and compressive strain $\epsilon_{zz} = -0.5\%$, as a function of QD length $L_z$ for $E_x = 5\, \text V / \mu$m; the analytical results of the corresponding quantities are shown in dashed lines. Here the QD length is defined as $L_z = (m^* \gamma_1/m_e)^{1/4}l_z \approx l_z$, where $m_e/\gamma_1$ is the averaged hole mass with $m_e$ being the electron mass and $\gamma_1$ is a Luttinger parameter~\cite{winkler2003spin,smref}. (b) ZFSs as a function of electric field $E_x$ for $L_z = 12\,$nm; inset: zoom at small electric fields, where the main anisotropy axis changes from the wire axis ($z$) to the SOI axis ($y$).}
\label{fig:ZFS_Lz_Ex}
\end{figure*}

\paragraph*{ Analytical theory. }

Large SOI emerges naturally in hole spin qubits encoded in long quantum dots, where the confinement potential in two directions is stronger than in the third one. Such nanostructures include a wide range of common spin qubit architectures, such as Si FinFETs~\cite{maurand2016cmos,kloeffel2018direct,urdampilleta2019gatebased,bosco2020hole}, squeezed QDs in planar Ge~\cite{bosco2021squeezed}, and Si and Ge NWs~\cite{kloeffel2011strong,gao2020site,froning2021strong,adelsberger2022hole}. Their response is well-described by an effective 1D low-energy Hamiltonian acting only on a few subbands.
 
We now focus on a QD defined in a NW with a square cross-section of side $L$. By resorting to Schrieffer-Wolff perturbation theory~\cite{bravy2011schrieffer} discussed in detail in Sec.~\ref{sm:cubicSOI} of~\cite{smref}, we find the effective Hamiltonian acting on the lowest pair of subbands as
\begin{equation}
H_{1} =  \frac{p_z^2}{2m^*} +  v p_z \sigma^y  + v_3 p_z^3 \sigma^y + \frac{\hbar^2 \gamma_1}{2m^* l_z^4} z^2\, ,
\label{eq:H1}
\end{equation}
up to third order in the momentum $p_z$ in the long-direction. Here, $m^*$ is the effective mass, $v$ and $v_3$ are the linear and cubic SOI, respectively, and $\sigma^y$ is a Pauli matrix. The QD is defined by a harmonic potential parametrized by the length $l_z$ and modelling the smooth electrostatic confinement produced by metallic gates. Eq.~\eqref{eq:H1} is valid when $l_z\gtrsim L/\pi$. Two holes confined in the same QD are described by the Hamiltonian $H_2 = H_1^{(1)} + H_1^{(2)} + V_c^{(1,2)}$, where $V_c^{(1,2)}$ is the effective Coulomb potential in the lowest subband sector. Coulomb interactions with higher subbands are negligible when $L/\pi < a_B$, where $a_B= 4\pi \epsilon_r \hbar^2/m^* e^2$ is the effective Bohr radius with $\epsilon_r$ being the dielectric constant of the material. The Coulomb potential $V_c^{(1,2)}$ is sketched in Fig.~\ref{fig:Effpot_splittings}(a), and is discussed in~\cite{smref}.

The linear SOI $v$ in Eq.~\eqref{eq:H1} can be eliminated exactly by a spin-dependent shift of momentum that leaves the potential unchanged, and only negligibly renormalizes the effective mass $m^*$~\cite{smref}. The two-particle Hamiltonian is then given by
\begin{equation}
\begin{split}
H_2 =& \frac{1}{4 m^*} P^2 + \frac{\hbar^2}{m^* l_z^4} Z^2 + \frac{1}{m^*}p^2  +\frac{\hbar^2}{4m^* l_z^4} z^2 +  V_c (z) \\
 &+ \mathcal P_3^+(\sigma_1^y + \sigma_2^y) + \mathcal P_3^- (\sigma_1^y - \sigma_2^y)\, ,
\label{eq:H2}
\end{split}
\end{equation}
where $Z = (z_1 + z_2)/2$ is the  center-of-mass (COM) coordinate with conjugate momentum $P = p_{z_1} + p_{z_2}$, and $z = z_1 - z_2$ is the relative coordinate with momentum $p = (p_{z_1} - p_{z_2})/2$. The cubic SOI  yields the perturbative corrections $\mathcal P_3^+=v_3 \left(\frac 1 8 P^3 + \frac 3 2 P p^2\right)$, and  $\mathcal P_3^-= v_3 \left(\frac 3 4 P^2 p + p^3\right)$ in the second line of Eq.~\eqref{eq:H2}; these terms mix relative and COM coordinates and are crucial for the ZFS.

At $v_3=0$, the Hamiltonian of the COM coordinates is a harmonic oscillator with an orbital energy $\Delta_o = \hbar^2/m^* l_z^2$, while the Hamiltonian of the relative coordinates is $H_\text{rel} = p^2/m^*  +\hbar^2 z^2 / 4m^* l_z^4 +  V_c (z)$. In a NW with a square cross-section and when $l_z\gtrsim a_B$, the effective 1D Coulomb interaction is well-approximated by $V_c(z)\approx \Delta_o [z^2 + (L/4)^2]^{-1/2} l^2_z / a_B$, where $L/4$ is a short-range cutoff of the potential derived in Sec.~\ref{sm:pmat} of~\cite{smref}. In this case, the system is fully described by two relative length scales $l_z/a_B$ and $L/ a_B$. Because the effective potential in $H_\text{rel}$ is an even function of $z$, the corresponding eigenfunctions have either even or odd parity, enabling the distinction between singlets (even) and triplets (odd) states. 

While in this work we focus on a single QD occupied by two holes, we emphasize that our theory is also valid for two tunnel-coupled QDs, the basic components of current spin-based quantum processors~\cite{hendrickx2020fast,hendrickx2020four}.
In fact, as sketched in  Fig.~\ref{fig:Effpot_splittings}(a), in a doubly occupied long QD, with $l_z \gtrsim a_B$, the Coulomb repulsion forces the two particles towards opposite ends of the dot~\cite{taut1993two,gao2020site,abadillo2021wigner}, effectively resulting in two coupled dots. 
We also remark that because $a_B\! \sim\! 12\,$nm ($a_B\! \sim\! 3\,$nm) in Ge (Si), the condition $l_z \gtrsim a_B$ of long QDs is typically respected in current experimental setups~\cite{froning2018single,katsaros2020zero,wang2022ultrafast}.

By a second order Schrieffer-Wolff transformation~\cite{bravy2011schrieffer} and projecting the two-particle Hamiltonian onto the lowest energy singlet and triplet states, we find that the exchange Hamiltonian is
\begin{equation}
\begin{split}
H_\text{eff} =& \frac 1 4 (\ex+\zfsd)\boldsymbol \sigma_1 \cdot \boldsymbol \sigma_2 - \frac 1 2 \zfsd \sigma^y_1 \sigma^y_2\\
&+ \frac 1 2 \boldsymbol \Delta^\perp\! \cdot (\boldsymbol \sigma_1^\perp\!+\!\boldsymbol \sigma_2^\perp) + \frac 1 2 \Delta^y (\sigma_1^y\!+\! \sigma_2^y)\, ,
\end{split}
\label{eq:Heff}
\end{equation}
where $\Delta^y$ is the Zeeman field parallel to the SOI, while ${\bf \Delta}^\perp = (\Delta^x, \Delta^z)$ are components perpendicular to it. 
The exchange splitting $\ex = \varepsilon^{}_{T_\pm} - \varepsilon^{}_S>0$ only weakly depends on $v_3$ and it is well approximated by  $\ex_0 =  \zeta\, \hbar^2 a_B^2/m^* l_z^4$, the energy gap between the lowest odd and even eigenstates of the relative coordinate Hamiltonian. We introduce the dimensionless coefficient $\zeta \sim 0.3 - 1$ for $0.8 < L/a_B <2$ and $a_B \lesssim l_z$~\cite{smref}.

Without magnetic fields,  $\Delta^i=0$ and Eq.~\eqref{eq:Heff} corresponds to an exchange Hamiltonian with a uni-axial anisotropy, i.e., $J_{xx} = J_{zz} = \ex$ and the anisotropy axis is aligned to the SOI (i.e., $y$-direction) with $J_{yy} = \ex + \zfsd$. As sketched in Fig.~\ref{fig:Effpot_splittings}(a), the ZFS $\zfsd$ lifts the degeneracy of the triplets $T_\pm$ and $T_0$, where the three triplets $T_{\pm,0}$ are defined with quantization axis along $y$-direction. From perturbation theory, we obtain~\cite{smref}
\begin{eqnarray}
\begin{split}
\zfsd =m^* v_3^2 \frac{\hbar^4}{l_z^4}  \eta\, .
\end{split}
\label{eq:ZFSD}
\end{eqnarray}
Here the dimensionless coefficient $\eta \sim 0.4 - 0.8$ includes  various combinations of dimensionless momentum matrix elements. The exact functional dependence of $\eta$ and $\zeta$ on $L$ and $l_z$ is discussed in detail in Sec.~\ref{sm:pmat} of~\cite{smref}. Because $\eta$ depends only weakly on the relative length scales $l_z/a_B$ and $L/ a_B$ in long QDs, to good approximation we find that $\zfsd\propto l_z^{-4}$. We also emphasize that this ZFS is strongly dependent on the cubic SOI and it requires a sizeable value of $v_3$, achievable only in hole QDs. The relative anisotropy of the exchange interactions is
\begin{equation}
\frac \zfsd \ex = \frac{m^{*2} v_3^2 \hbar^2}{a_B^2} \frac \eta \zeta\, ,
\end{equation}
where $\eta/\zeta \sim 1-5$ depends weakly on $a_B$ and therefore, the anisotropy scales as $\zfsd/\ex \propto (m^*)^4$.

The magnetic field dependence of the triplet states can also be deduced straightforwardly from Eq.~\eqref{eq:Heff} and it is sketched in  Fig~\ref{fig:Effpot_splittings}(b). If the magnetic field is applied parallel to the SOI (i.e. the anisotropy axis) the non-degenerate triplet $T_0$ is unaffected by the field and $\varepsilon^{}_{T_0} = \ex+\zfsd$, whereas the degenerate triplets $T_\pm$ split linearly with the Zeeman field as $\varepsilon_{T_\pm}^{} = \ex\pm\Delta^y$. In contrast, if the field is applied perpendicular to the SOI, one of the degenerate triplets, e.g., $T'_0$, stays at the same energy $\varepsilon_{T'_0}^{} = \ex$, while the remaining triplets $T'_\pm$ split quadratically as $\varepsilon_{T'_\pm}^{} = \ex + \zfsd/2  \pm \sqrt{ \zfsd^2/4+|\Delta^\perp|^2} $ at small Zeeman fields. This signature of the exchange anisotropy is consistent with recent experimental observations in Ref.~\citep{katsaros2020zero}, supporting our theory of ZFS in Ge hut wires.

\paragraph*{ Numerics. } 

We confirm our analytical results by comparing them with a numerical simulation of long QDs in square Ge and Si NWs with side length $L$ based on the 6-band Kane model~\cite{winkler2003spin}. By imposing hard-wall boundary conditions at the edge of the NW cross-section, we obtain an effective 1D model including several transversal subbands. With a third order Schrieffer-Wolff transformation, we then fold the higher energy subbands down to the lowest four subbands, also accounting for terms that are cubic in momentum. We emphasize that in contrast to our analytical treatment, where we only account for a single pair of subbands, see Eq.~\eqref{eq:H1}, our numerical treatment also includes a pair of higher-energy subbands~\cite{smref}. Furthermore, we include Coulomb interaction matrix elements that couple different subbands, as well as short-range interband corrections to the Coulomb interaction~\cite{secchi2020inter}, that we identify as an alternative source of ZFS. In our simulation, we also consider a compressive strain along the wire, with $\epsilon_{zz} = -0.5\%$, ensuring that the lowest band has a positive effective mass~\cite{kloeffel2011strong,kloeffel2018direct}. More details on the numerical simulation are provided in Sec.~\ref{sm:numcalc} of~\cite{smref}, where we also confirm the validity of our four subband model by comparing it to a full three-dimensional simulation.

In Fig.~\ref{fig:ZFS_Lz_Ex}(a), we compare the numerical simulation of a Ge NW with $L = 10\,$nm with the analytical formulas of the exchange splitting $\ex$ and the ZFS in Eq.~\eqref{eq:ZFSD} as a function of QD length $L_z$. In this calculation, the $\{x,y,z\}$ axes coincide with the $\braket{100}$ crystallographic directions. Strikingly, the numerical exchange splitting $\ex$ is in excellent agreement with the analytical formula, and also  $\zfsd$ is reasonably well captured by  the simple Eq.~\eqref{eq:ZFSD} in a wide range of QD sizes. We emphasize that due to the weak dependence of the coefficient $\eta$ on the side length $L$ in long QDs ($L,a_B<l_z$) Eq.~\eqref{eq:ZFSD} can accurately estimate the ZFS in general architectures.

The numerical solution in Fig.~\ref{fig:ZFS_Lz_Ex}(a) also reveals an additional ZFS of the remaining two triplet states, that emerges because of the short-range corrections to the Coulomb interaction~\cite{secchi2020inter}. These corrections stem from the atomistic interactions of the $p$-type Bloch functions and induce mixing between the different bulk hole bands. The contribution of the short-range corrections to the effective Hamiltonian of Eq.~\eqref{eq:Heff} can be written as
\begin{equation}
\begin{split}
H_\text{eff,\,s-r} = \frac 1 2 \zfse \sigma^z_1 \sigma^z_2\, ,
\end{split}
\label{eq:Heffsr}
\end{equation}
where $\zfse$ is the exchange anisotropy along the NW ($z$-direction). This ZFS induces an energy gap $\zfse$ between the triplets $\ket{T_0}$, $\ket{T_a} = (\ket{T_+} + \ket{T_-})/\sqrt 2$, and the remaining states [the singlet $\ket S$ and the third triplet $\ket{T_b} = (\ket{T_+} - \ket{T_-})/\sqrt 2$], thereby lifting the remaining triplet degeneracy at zero magnetic field.

The exchange anisotropy $\zfse$ induced by the short-range Coulomb interaction is also present without external electric fields, where the SOI vanishes [see Fig.~\ref{fig:ZFS_Lz_Ex}(b)]. In this special case because of the fourfold symmetry of the system, the anisotropy axis is aligned to the wire~\cite{solyom2007fundamentals,koster1963properties,smref}. If an electric field is applied perpendicular to the wire, the symmetry is reduced and the remaining degeneracy is also lifted. (For a detailed symmetry analysis of different wire geometries see Sec.~\ref{sm:tripletdeg} of \cite{smref}.) At small $E_x$, the ZFS $\zfsd$ increases quadratically with the electric field, because $v_3 \sim E_x$, and eventually overcomes $\zfse$ [see the inset in Fig.~\ref{fig:ZFS_Lz_Ex}(b)], aligning the main anisotropy axis to the SOI. For higher electric fields, $v_3$ (and thus $\zfsd$) reaches a maximum value and starts to decrease, in analogy to the linear SOI $v$ in various NW geometries~\cite{kloeffel2018direct,bosco2020hole}.

\begin{figure}
\centering
\includegraphics[width=0.9\columnwidth]{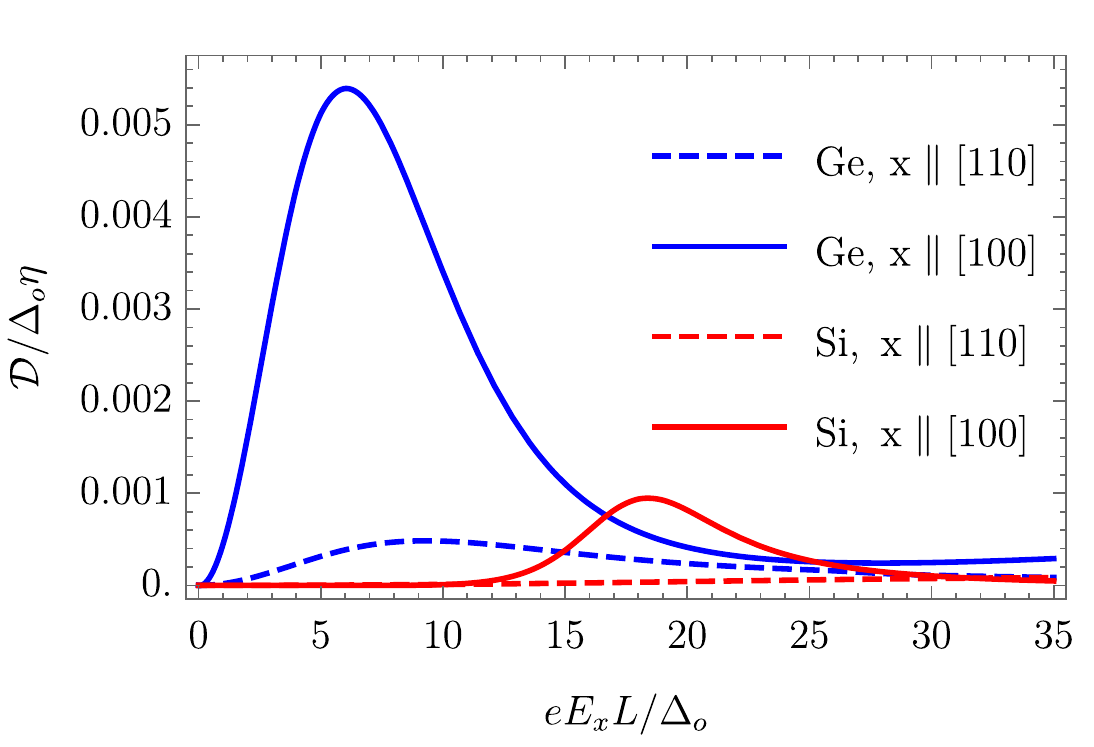}
\caption{Dependence of the ZFS $\zfsd$ in Eq. \eqref{eq:ZFSD}  on the electric field $E_x$. With blue (red) lines, we show Ge (Si) for two different growth directions and split-off gap  $\Delta_{SO} \sim 150 \Delta_o$ ($\Delta_{SO} \sim 4 \Delta_o$).  Here, we consider $l_z = L=2a_B$, $z\parallel [001]$, and we use the strain $\epsilon_{zz} = -0.5\%$. The orbital energy is $\Delta_o = \hbar^2 /m^* l_z^2$.}
\label{fig:ZFS_Ge_Si}
\end{figure}

The electric field dependence of the ZFS in Eq.~\eqref{eq:ZFSD} is dominated by $v_3^2$ and therefore $\zfsd$ is highly tunable by the external gate potentials and by the QD design. In particular,  in Fig.~\ref{fig:ZFS_Ge_Si} we show  $\zfsd$  as a function of electric field in Ge and Si NWs for different growth directions. For both growth directions, the ZFS --relative to the orbital splitting-- is significantly smaller in Si than in Ge. This reduction is a result of the hybridization of HHs and LHs with the spin-orbit split-off band that is much closer in Si ($\Delta_{SO} = 44\,$meV) than in Ge ($\Delta_{SO} = 296\,$meV)~\cite{winkler2003spin}, effectively decreasing the HH-LH mixing and the SOI~\cite{bosco2020hole}.

The ZFS also varies substantially between different growth directions for both materials as shown in Fig.~\ref{fig:ZFS_Ge_Si}. The strong dependence of the SOI on the growth direction is well-known in Si nanowires~\cite{kloeffel2018direct,bosco2020hole}, and it is also significant in Ge. Strikingly, the linear SOI $v$ changes only slightly in Ge between the two growth directions~\cite{kloeffel2018direct,bosco2021squeezed}, but the cubic SOI $v_3$ is strongly altered between the two cases, yielding an order of magnitude larger ZFS  when $x\parallel[110]$. This enhancement can be explained by considering that the cubic SOI is a higher order correction that involves more subbands,  making $v_3$  more sensitive to the growth direction and to the design of the QD. 
This finding stresses once again that the ZFS in hole QDs is induced by the cubic SOI $v_3$ and that there is no direct relation between the ZFS and the linear SOI $v$.

\paragraph*{ Conclusions. } 
We presented a simple analytical model  explaining the large anomalous triplet splitting at zero magnetic field, emerging in QDs occupied by two holes and shedding some light on recent experimental findings~\cite{katsaros2020zero}. 
We related the ZFS to a cubic SOI that is externally tunable by electric fields and can be engineered by the design of the QD. In striking contrast to linear SOI effects, the ZFS is found to depend significantly on the growth direction not only in  Si but also in Ge QDs, where such anisotropic effects are typically small~\cite{kloeffel2011strong,kloeffel2018direct}. The SOI induced ZFS is also found to be orders of magnitude larger than short-range corrections to the Coulomb interaction, an alternative mechanisms for the ZFS of triplet states. While our analytical model focuses on doubly occupied long QDs, our findings are also valid in two tunnel-coupled QDs, the main building blocks of current spin-based quantum processors, and thus our work has deep implications for the design of future scalable quantum computing architectures with hole spin qubits. \\

We thank A. P{\'a}lyi, D. Miserev, and G. Katsaros for the fruitful discussions. 
This work was supported as a part of NCCR SPIN funded by the Swiss National Science Foundation (grant number 51NF40-180604).

\bibliography{references}

\pagebreak

\clearpage
\onecolumngrid
\vspace*{0.1cm}
\begin{center}
	\large{\bf Supplementary Material to \\`Anomalous zero-field splitting  for hole spin qubits in Si and Ge quantum dots'\\}
\end{center}
\begin{center}
	Bence Het\'enyi, Stefano Bosco, and Daniel Loss \\
	{\it\small Department of Physics, University of Basel, Klingelbergstrasse 82, CH-4056 Basel, Switzerland}\\
	(Dated: \today)
\end{center}
\vspace*{0.1cm}

\onecolumngrid
\begin{center}
\parbox{400pt}{\small \hspace{5pt} Here we provide an explicit derivation of the zero-field splitting formula shown in the main text and reveal further details of the numerical calculation. A symmetry analysis of the triplet degeneracy is also included, as well as the exact form of the interband Coulomb corrections. Furthermore we compare the numerical calculation for long QDs with a different numerical approach working for short QDs. We find a good agreement of the two calculations confirming the results presented in the main text. }
\end{center}

\setcounter{equation}{0}
\setcounter{figure}{0}
\setcounter{table}{0}
\setcounter{section}{0}

\renewcommand{\theequation}{S\arabic{equation}}
\renewcommand{\thefigure}{S\arabic{figure}}
\renewcommand{\thesection}{S\arabic{section}}
\renewcommand{\bibnumfmt}[1]{[S#1]}
\renewcommand{\citenumfont}[1]{S#1}

\section{Zero-field splitting induced by cubic spin-orbit interaction}
\label{sm:cubicSOI}

Here, we discuss in more detail the effective model of the zero-field splitting introduced in the main text. When the QD is elongated in the $z$ direction, the low-energy behaviour of the system can be described by an effective model where only the lowest subbands of a quasi-1D system are taken into account. Here we present a two-band minimal model that is sufficient to explain the mechanism. This model gives a rather accurate estimate of the zero-field splitting in a wide range of cases. We consider the Hamiltonian up to third order in momentum including a harmonic confinement
\begin{equation}
H_1 =  \frac{p_z^2}{2m} + v p_z \sigma^y + v_3 p_z^3 \sigma^y + \frac{\hbar^2\gamma_1}{2m_e L_z^4}z^2 \, ,
\label{smeq:HNWSOI}
\end{equation}
where $m$ is the effective mass, $v$ is the spin-orbit velocity, $v_3$ is the coefficient of the SOI cubic in momentum $p_z$, and $L_z$ the harmonic confinement length of the QD. Note that the linear  and the cubic SOI terms need to be aligned to the same SOI axis (here $\sigma^y$), otherwise one could construct second order terms at $B = 0$ such as $\braket {p^3}_{mn}\braket {p}_{nm} \sigma^x \sigma^y \sim p^4 \sigma^z$ that would break time-reversal symmetry.

We apply a unitary transformation $U(p_0) = \exp (-ip_0z\sigma^y/\hbar)$ on the Hamiltonian in Eq.~\eqref{smeq:HNWSOI} that shifts the momentum as $p_z \rightarrow p_z-p_0\sigma^y$. By choosing the momentum shift $p_0= (1-\sqrt{1-12m^2 v_3 v})/6 m v_3$ such that the terms linear in $p_z$ vanish, we obtain the Hamiltonian
\begin{equation}
\tilde H_1  = U^\dagger(p_0) H_1 U(p_0) =  \frac{p_z^2}{2m^*} + v_3 p_z^3 \sigma^y + \frac{\hbar^2\gamma_1}{2m_e L_z^4}z^2\, , \text{ with }  \hspace{0.2cm} \frac{1}{m^*} = \frac{1}{m}\sqrt{1-12m^2 v_3 v}\, ,
\label{smeq:HNW}
\end{equation}
where $12m^2 v_3 v \ll 1$ even for strong electric fields and we introduce the renormalized harmonic confinement length as $l_z = (m_e/m^*\gamma_1)^{1/4}L_z$. In the followings we omit the tilde from the transformed Hamiltonian $\tilde H_1$ (as in the main text).

We now consider two-particle systems and we include Coulomb interaction in the 1D Hamiltonian of Eq.~\eqref{smeq:HNW}. Then the two-particle Hamiltonian reads
\begin{equation}
H_2 = H_1^{(1)}+ H_1^{(2)}+  \frac{\hbar^2}{2m^* l_z^4}(z_1^2 + z_2^2) + V_c (z_1 - z_2)\, .
\end{equation}
Here, $V_c (z_1 - z_2)$ is the effective 1D Coulomb interaction obtained by projecting the Coulomb interaction onto the lowest subband with the corresponding lowest eigenstates of the full 3D Hamiltonian at $p_z = 0$. Due to this projection, the singularity of the Coulomb interaction is cut off in $V_c (z_1 - z_2)$ at a distance $|z_1 - z_2| \sim L\ll l_z$ determined by the transversal confinement (see Sec.~\ref{sm:pmat} for a fitting formula at square cross section). Moving to the center-of-mass (COM) frame one obtains, 
\begin{equation}
H_2 = \frac{1}{4 m^*} P^2 +  \frac{\hbar^2}{m^* l_z^4} Z^2 + \frac{p^2}{m^*}  + \frac{\hbar^2}{4m^* l_z^4} z^2 +  V_c (z) + v_3 \left(\frac 1 8 P^3 + \frac 3 2 P p^2\right) (\sigma_1^y + \sigma_2^y) + v_3 \left(\frac 3 4 P^2 p + p^3\right) (\sigma_1^y - \sigma_2^y)\, ,
\label{smeq:H2COM}
\end{equation}
where the position and the conjugate momentum for the COM and the relative coordinates read 
\begin{subequations}
\begin{equation}
Z = (z_1 + z_2)/2\, , \hspace{0.5cm} P = p_{z_1} + p_{z_2}\, ,
\end{equation}
\begin{equation}
z = z_1 - z_2\, , \hspace{0.5cm} p = (p_{z_1} - p_{z_2})/2\, ,
\end{equation}
\label{smeq:COMcoord}
\end{subequations}
respectively. Since the cubic SOI term $\propto v_3$ is obtained by a third order Schrieffer-Wolff (SW) transformation, it is suppressed by the subband gap compared to other terms of the Hamiltonian. If the subband gap is large compared to $v_3 \hbar^3/l_z^3$, the cubic SOI term can be treated as a small perturbation that couples both the COM and relative coordinates with the spin degree of freedom. We divide Eq.~\eqref{smeq:H2COM} into three terms
\begin{subequations}
\begin{equation}
H^\text{\tiny{COM}}_2 = \frac{1}{4 m^*} P^2 + \frac{\hbar^2}{m^* l_z^4} Z^2\, ,
\label{smeq:Hcom}
\end{equation}
\begin{equation}
H^\text{rel}_2 =\frac{p^2}{m^*}  + \frac{\hbar^2}{4m^* l_z^4} z^2 +  V_c (z)\, ,
\label{smeq:Hrel}
\end{equation}
\begin{eqnarray}
\begin{split}
V &=v_3 \left(\frac 1 8 P^3 + \frac 3 2 P p^2\right) (\sigma_1^y + \sigma_2^y) + v_3 \left(\frac 3 4 P^2 p + p^3\right) (\sigma_1^y - \sigma_2^y) \equiv \mathcal P_3^+(\sigma_1^y + \sigma_2^y) +\mathcal P_3^- (\sigma_1^y - \sigma_2^y)\, .
\label{smeq:Vpm}
\end{split}
\end{eqnarray}
\end{subequations}

The COM Hamiltonian of Eq.~\eqref{smeq:Hcom} can be rewritten using the harmonic oscillator ladder operators defined as $P = i(a^\dagger - a)\hbar/l_z$ and $Z = (a^\dagger + a)l_z/2$ such that $ H^\text{\tiny{COM}}_2 = \Delta_o a^\dagger a$, where $\Delta_o = \hbar^2/m^* l_z^2$ is the energy splitting of the COM mode. In contrast to $H^\text{\tiny{COM}}$ the Hamiltonian $H^\text{rel}$ cannot be diagonalized exactly. Nevertheless exploiting the $z\leftrightarrow -z$ symmetry of the Hamiltonian, we can denote the lowest even (odd) eigenstate with $S$ ($T$) referring to their singlet-like (triplet-like) behaviour under particle exchange. Even though the 1D two-particle problem of a harmonic potential in the long QD limit ($l_z \gg a_B$) can be treated analytically in the Hund-Mulliken approximation~\cite{taut1993sm,gao2020sm}, here we resort to the numerical solution of this problem because we are interested in the $l_z\gtrsim a_B$ regime where this approximation is not accurate.

By using a second order Schrieffer-Wolff transformation, we project the Hamiltonian to the ground state of the COM Hamiltonian and the two energetically lowest  eigenstates of the relative coordinate Hamiltonian (i.e., one singlet-like and one triplet-like state). Thereby, an effective low-energy Hamiltonian is obtained, from which the anisotropy axis can be deduced and the magnetic field dependence can be straightforwardly discussed. The effective low-energy Hamiltonian reads
\begin{equation}
H_\text{eff} = -\ex_0 \ket{\chi^{}_S} \bra{\chi^{}_S}+ W_\text{eff}\, ,
\label{smeq:Heff}
\end{equation}
where $\ex_0$ is the energy splitting between the lowest-energy eigenstates of \eqref{smeq:Hrel} and $\ket{\chi_S} = (\ket{\uparrow}_1 \ket{\downarrow}_2 - \ket{\downarrow}_1 \ket{\uparrow}_2)/\sqrt{2}$ is the spin part of the singlet wavefunction, and where we choose the spin quantization axis along $y$-direction, i.e.  $\sigma_i^y \ket{\uparrow\, (\downarrow)}_i = \pm \ket{\uparrow\, (\downarrow)}_i$. In the following we also need the three triplet-like states, denoted by  $\ket{\chi_{T_0}} = (\ket{\uparrow}_1 \ket{\downarrow}_2 + \ket{\downarrow}_1 \ket{\uparrow}_2)/\sqrt{2}$,  
$\ket{\chi_{T_+}} = \ket{\uparrow}_1 \ket{\uparrow}_2 $, and $\ket{\chi_{T_-}} = \ket{\downarrow}_1 \ket{\downarrow}_2 $. 
The effective coupling 
\begin{eqnarray}
\begin{split}
W_\text{eff} =& - \frac i {2 \hbar} \lim_{\eta \rightarrow 0^+} \int \limits_0^\infty \! dt\, e^{-\eta t} \braket{\left[ V(t), V \right]}\\
=&  - \frac i {\hbar} \lim_{\eta \rightarrow 0^+} \int \limits_0^\infty \! dt\, e^{-\eta t} \left \{ \braket{[ \mathcal P_3^+(t), \mathcal P_3^+ ]} (1 + \sigma_1^y \sigma_2^y )  + \braket{[ \mathcal P_3^-(t), \mathcal P_3^- ]}(1 - \sigma_1^y \sigma_2^y ) \right \} \, ,
\label{smeq:Weffnopauli}
\end{split}
\end{eqnarray}
stems from the cubic SOI terms of Eq.~\eqref{smeq:Vpm}, where $V(t) = e^{iH_0t/\hbar} V e^{-iH_0t/\hbar}$ is the perturbation in the interaction picture, with the unperturbed Hamiltonian $H_0 =  H^\text{\tiny{COM}}_2 + H^\text{rel}_2$. Also, the expectation values in Eq.~\eqref{smeq:Weffnopauli} project the effective Hamiltonian onto the low-energy singlet-triplet subspace, and in the second equation we exploited the fact that $(\sigma_1^y + \sigma_2^y)(\sigma_1^y - \sigma_2^y) = 0$. 

To include Pauli's principle, we restrict the Hilbert space to the antisymmetric 2-particle solutions by projecting Eq.~\eqref{smeq:Weffnopauli} onto the lowest-energy singlet and triplet basis. The respective spin matrices projected onto the triplet sector can be written as
\begin{subequations}
\begin{equation}
(1+\sigma_1^y\sigma_2^y)^{}_{T} = 2 (\ket{\chi^{}_{T_+}} \bra{\chi^{}_{T_+}} + \ket{\chi^{}_{T_-}} \bra{\chi^{}_{T_-}}) =  1+\sigma_1^y\sigma_2^y\, ,
\end{equation}
\begin{equation}
(1-\sigma_1^y\sigma_2^y)^{}_{T} = 2 \ket{\chi^{}_{T_0}} \bra{\chi^{}_{T_0}} = \frac 1 2 - \sigma_1^y\sigma_2^y + \frac 1 2 \boldsymbol \sigma_1 \cdot \boldsymbol \sigma_2\, ,
\end{equation}
while the corresponding projection of the singlets reads 
\begin{equation}
(1+\sigma_1^y\sigma_2^y)^{}_{S} = 0\, ,
\end{equation}
\begin{equation}
(1-\sigma_1^y\sigma_2^y)^{}_{S} = 2\ket{\chi^{}_S} \bra{\chi^{}_S} = \frac 1 2 - \frac 1 2 \boldsymbol \sigma_1 \cdot \boldsymbol \sigma_2\, .
\end{equation}
\end{subequations}
Exploiting that the spin parts of the effective low-energy Hamiltonian do not couple the singlet with the triplet sectors one may write the perturbation as
\begin{eqnarray}
W_\text{eff} =  W^+_T (1 + \sigma_1^y \sigma_2^y ) + W^-_T \left(\frac 1 2 - \sigma_1^y\sigma_2^y + \frac 1 2 \boldsymbol \sigma_1 \cdot \boldsymbol \sigma_2\right) + W^-_S \left(\frac 1 2 - \frac 1 2 \boldsymbol \sigma_1 \cdot \boldsymbol \sigma_2\right) \, ,
\label{smeq:Weff}
\end{eqnarray}
where the prefactors, in analogy with Eq.~\eqref{smeq:Weffnopauli} are given by
\begin{equation}
W^\pm_{S(T)} = - \frac i {\hbar} \lim_{\eta \rightarrow 0^+} \int \limits_0^\infty \! dt\, e^{-\eta t} \braket{[ \mathcal P_3^\pm(t), \mathcal P_3^\pm] }_{S(T)}.
\label{smeq:WpmST}
\end{equation}
Here, $\braket{\dots}_{S(T)}$ is the expectation value taken with respect to the state $\ket{0,\psi_{S(T)}} = \ket{0}\ket{\psi_{S(T)}}$, where $\ket{0}$ is the ground state of the COM Hamiltonian and $\ket{\psi_{S(T)}}$ is the lowest-energy singlet-like (triplet-like) eigenstate of the relative coordinate Hamiltonian in Eq.~\eqref{smeq:Hrel}.

Substituting the effective coupling \eqref{smeq:Weff} into \eqref{smeq:Heff}, the effective Hamiltonian can be written in the following form
\begin{equation}
H_\text{eff} = \frac 1 4 (\ex+\zfsd)\boldsymbol \sigma_1 \cdot \boldsymbol \sigma_2 - \frac 1 2 \zfsd \sigma^y_1 \sigma^y_2\, ,
\end{equation}
where $\zfsd = 2(W^-_T - W^+_T)$ is the exchange anisotropy responsible for the zero-field splitting and $\ex = \ex_0 + 2(W^+_T-W^-_S)$ is the exchange splitting between the singlet and the $T_\pm$ doublet. In order to determine the zero-field splitting $\zfsd$ we need to calculate the quantities $W^\pm_{S(T)}$. To this aim, we first write the time-evolution of the COM momentum as 
\begin{equation}
P(t) = i\frac{\hbar}{l_z} \left(a^\dagger e^{i\Delta_o t/\hbar} - a e^{-i\Delta_o t/\hbar}\right)\, ,
\label{smeq:Pt}
\end{equation}
while higher powers of the momentum can be expressed straightforwardly by using the creation and annihilation operators $a^\dagger$ and $a$. For the matrix elements of the relative momentum we can only exploit the even/odd parity of the basis states to write the matrix elements of $p$ and $p^3$ between $\ket{\psi_{S(T)}}$ and an arbitrary state $\ket{\psi_n}$ as
\begin{subequations}
\begin{equation}
\bra{\psi_S} p^{1,3}(t) \ket{\psi_n} = \sum_{i} \delta_{n,T_i} \braket{p^{1,3}}_{S,T_i} e^{-i(\varepsilon_{T_i} -\varepsilon_{S}) t/\hbar}\, ,
\label{smeq:pSt}
\end{equation}
\begin{equation}
\bra{\psi_T} p^{1,3}(t) \ket{\psi_n} = \sum_{i} \delta_{n,S_i} \braket{p^{1,3}}_{T,S_i} e^{-i(\varepsilon_{S_i} -\varepsilon_{T}) t/\hbar}\, ,
\label{smeq:pTt}
\end{equation}
\end{subequations}
where $S_i$ ($T_i$) denote the higher energy even (odd) states for $i = 1,2,3\dots$ The matrix elements of $p^2(t)$ can be written analogously and only couple even (odd) states to higher even (odd) states. In the next step the projected commutators in Eq.~\eqref{smeq:WpmST} are obtained using Eqs.~\eqref{smeq:Pt}-\eqref{smeq:pTt}, resulting in
\begin{subequations}
\begin{eqnarray}
\begin{split}
\braket{[ \mathcal P_3^+(t), \mathcal P_3^+ ]}_T =& \frac{9}{64} v_3^2\frac{\hbar^6}{l_z^6} e^{-i \Delta_o t/\hbar}  + \frac{6}{64} v_3^2\frac{\hbar^6}{l_z^6} e^{-3i \Delta_o t/\hbar} + \frac{9}{8} v_3^2\frac{\hbar^4}{l_z^4} \braket{p^2}_{_{TT}} e^{-i \Delta_o t/\hbar} \\
&+\frac{9}{4} v_3^2\frac{\hbar^2}{l_z^2} \sum_i |\braket{p^2}_{_{TT_i}}|^2 e^{-i (\Delta_o+ \varepsilon_{T_i} - \varepsilon_T ) t/\hbar} - h.c.\, ,
\end{split}
\end{eqnarray}
\begin{eqnarray}
\begin{split}
\braket{[ \mathcal P_3^-(t), \mathcal P_3^- ]}_T =& \frac{3}{2} v_3^2\frac{\hbar^2}{l_z^2} \sum_i \text{Re}[ \braket{p^3}_{_{TS_i}} \braket{p}_{_{S_i T}} ] e^{-i (\varepsilon_{S_i} - \varepsilon_T) t/\hbar}  + v_3^2 \sum_i |\! \braket{p^3}_{_{TS_i}}\!|^2 e^{-i (\varepsilon_{S_i} - \varepsilon_T) t/\hbar} \\
& + \frac{9}{16} v_3^2 \frac{\hbar^4}{l_z^4}\sum_i  |\! \braket{p}_{_{TS_i}}\!|^2 e^{-i (\varepsilon_{S_i} - \varepsilon_T) t/\hbar} +\frac{9}{8} v_3^2 \frac{\hbar^4}{l_z^4}\sum_i  |\! \braket{p}_{_{TS_i}}\!|^2 e^{-i (2\Delta_o  + \varepsilon_{S_i} - \varepsilon_T) t/\hbar} 
 - h.c.\, ,
\end{split}
\end{eqnarray}
\begin{eqnarray}
\begin{split}
\braket{[ \mathcal P_3^-(t), \mathcal P_3^- ]}_S =& \frac{3}{2} v_3^2\frac{\hbar^2}{l_z^2} \sum_i \text{Re}[ \braket{p^3}_{_{ST_i}}\! \braket{p}_{_{T_i S}} ] e^{-i (\varepsilon_{T_i} - \varepsilon_S) t/\hbar}  + v_3^2 \sum_i |\! \braket{p^3}_{_{ST_i}}\!|^2 e^{-i (\varepsilon_{T_i} - \varepsilon_S) t/\hbar} \\
&+\frac{9}{16} v_3^2 \frac{\hbar^4}{l_z^4}\sum_i  |\! \braket{p}_{_{ST_i}}\!|^2 e^{-i (\varepsilon_{T_i} - \varepsilon_S) t/\hbar} +\frac{9}{8} v_3^2 \frac{\hbar^4}{l_z^4}\sum_i  |\! \braket{p}_{_{ST_i}}\!|^2 e^{-i (2\Delta_o  + \varepsilon_{T_i} - \varepsilon_S) t/\hbar} 
 - h.c.\, ,
\end{split}
\end{eqnarray}
\end{subequations}
where the commutator in $W^+_S$ is not listed since it does not contribute to the effective coupling in Eq.~\eqref{smeq:Weff}. The time integrals in Eq.~\eqref{smeq:WpmST} can be evaluated using $\int_0^\infty e^{i\omega t -0^+ t} = -i/(\omega-i0^+)$.

Finally, the zero-field splitting $\zfsd$ is expressed in terms of momentum matrix elements as
\begin{eqnarray}
\begin{split}
\zfsd =& \frac{11}{16} v_3^2\frac{\hbar^6}{l_z^6}\frac 1 {\Delta_o} + \frac{9}{2} v_3^2\frac{\hbar^4}{l_z^4} \frac{\braket{p^2}_{_{TT}}}{\Delta_o} + 9 v_3^2\frac{\hbar^2}{l_z^2} \sum_i  \frac{|\braket{p^2}_{_{TT_i}}|^2}{\Delta_o+ \varepsilon_{T_i} - \varepsilon_T } -6 v_3^2\frac{\hbar^2}{l_z^2} \sum_i  \frac{\text{Re}[ \braket{p^3}_{_{TS_i}}\! \braket{p}_{_{S_i T}} ]}{\varepsilon_{S_i} - \varepsilon_T}\\
&- 4 v_3^2 \sum_i \frac{|\! \braket{p^3}_{_{TS_i}}\!|^2 }{\varepsilon_{S_i} - \varepsilon_T}  -\frac{9}{4} v_3^2 \frac{\hbar^4}{l_z^4}\sum_i   \frac{|\! \braket{p}_{_{TS_i}}\!|^2}{\varepsilon_{S_i} - \varepsilon_T} -\frac{9}{2} v_3^2 \frac{\hbar^4}{l_z^4}\sum_i   \frac{ |\! \braket{p}_{_{TS_i}}\!|^2}{2\Delta_o  +\varepsilon_{S_i} - \varepsilon_T} \equiv m^* v_3^2 \frac{\hbar^4}{l_z^4} \eta \, ,
\end{split}
\label{smeq:ZFS}
\end{eqnarray}
where we defined the dimensionless prefactor $\eta$ as in Eq.~\eqref{eq:ZFSD} of the main text. We show its functional dependence in Sec.~\ref{sm:pmat}. Moreover, the exchange splitting is given by
\begin{eqnarray}
\begin{split}
\ex = \ex_0 &- \frac{11}{16} v_3^2\frac{\hbar^6}{l_z^6}\frac 1 {\Delta_o} - \frac{9}{2} v_3^2\frac{\hbar^4}{l_z^4} \frac{\braket{p^2}_{_{TT}}}{\Delta_o} - 9 v_3^2\frac{\hbar^2}{l_z^2} \sum_i  \frac{|\braket{p^2}_{_{TT_i}}|^2}{\Delta_o+ \varepsilon_{T_i} - \varepsilon_T } + 6 v_3^2\frac{\hbar^2}{l_z^2} \sum_i \frac{ \text{Re}[ \braket{p^3}_{_{ST_i}}\! \braket{p}_{_{T_i S}} ]}{\varepsilon_{T_i} - \varepsilon_S} \\
&+ 4 v_3^2 \sum_i  \frac{|\! \braket{p^3}_{_{ST_i}}\!|^2}{\varepsilon_{T_i} - \varepsilon_S} +\frac{9}{4} v_3^2 \frac{\hbar^4}{l_z^4}\sum_i  \frac{ |\! \braket{p}_{_{ST_i}}\!|^2 }{\varepsilon_{T_i} - \varepsilon_S} +\frac{9}{2} v_3^2 \frac{\hbar^4}{l_z^4}\sum_i   \frac{|\! \braket{p}_{_{ST_i}}\!|^2}{2\Delta_o  +\varepsilon_{T_i} - \varepsilon_S} \approx \ex_0 \, ,
\end{split}
\end{eqnarray}
where $\ex_0 = \varepsilon_T - \varepsilon_S$ is the triplet-singlet splitting of the unperturbed Hamiltonian. These equations correspond to the ones reported in the main text.

\subsection{Momentum matrix elements of the relative coordinate}
\label{sm:pmat}

We now provide more details on the magnitude of the exchange $\ex$ and of the zero-field splitting $\zfsd$. The analytical result of the ZFS in Eq.~\eqref{smeq:ZFS} involves a number of matrix elements of different powers of momentum between the eigenstates of the Hamiltonian $H_2^\text{rel}$ of the relative coordinate in Eq.~\eqref{smeq:Hrel}. Since the Hamiltonian contains the effective 1D potential $V_c(z_1 - z_2)$, it is difficult to estimate this matrix elements in general. 

\begin{figure}
\includegraphics[width=0.8\textwidth]{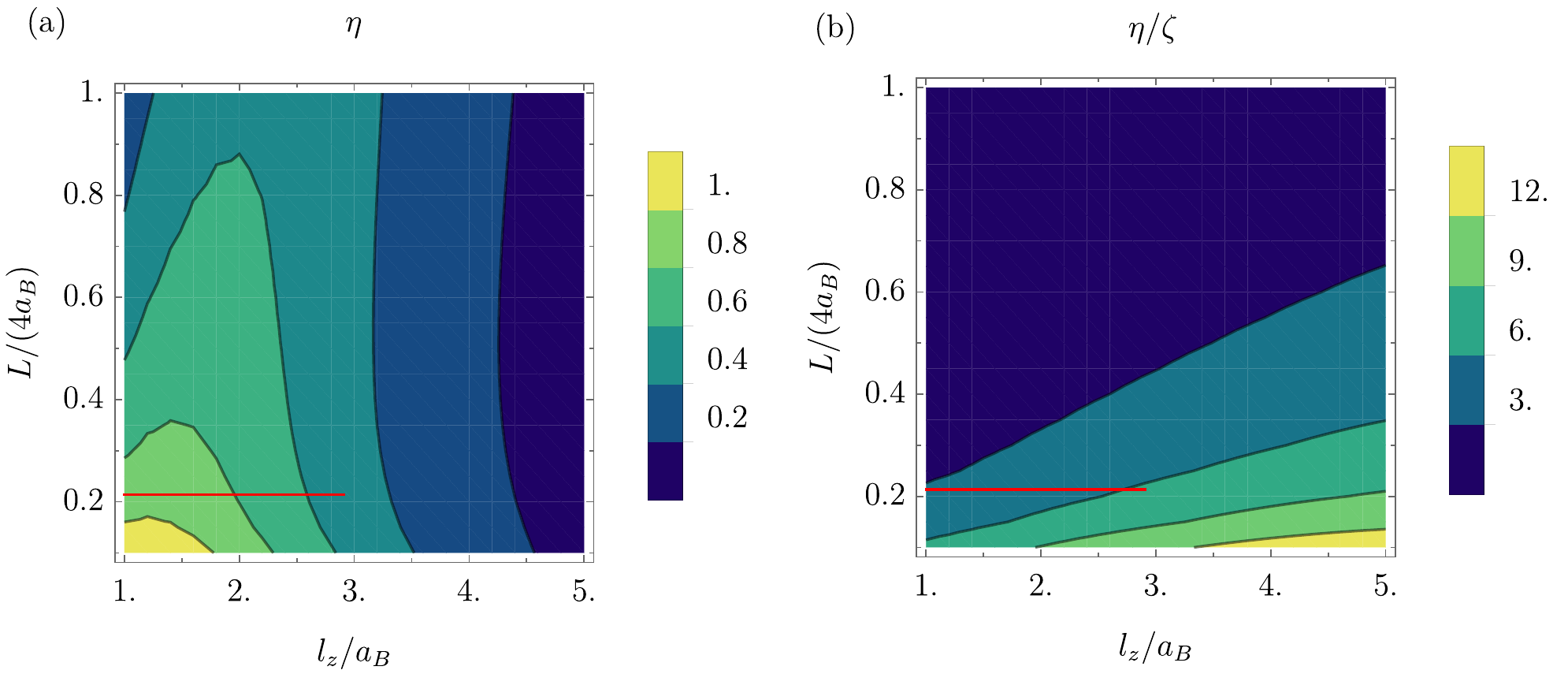}
\caption{(a) Coefficient $\eta$ of the zero-field splitting $\zfsd$ as a function of QD length $l_z$ and NW width $L$. (b) Ratio $\eta / \zeta$ of the anisotropy $\zfsd / \ex$ as a function of QD length and NW width. The thin red line corresponds to the region plotted in Fig.~\ref{fig:ZFS_Lz_Ex}(a) of the main text. For reference $a_B \approx 11.7\,$nm in Ge and $a_B \approx 2.7\,$nm in Si. } 
\label{smfig:ZFScoeff_function}
\end{figure}

Here we restrict our attention to nanowires with square cross section and side length of $L$ and calculate the effective 1D Coulomb potential numerically as discussed in Sec.~\ref{sm:numcalc}. We find that the relevant momentum matrix elements are very well reproduced by using the following effective potential
\begin{equation}
V_c (z) \approx \frac{e^2}{4\pi \epsilon_r} \frac{1}{\sqrt{z^2+(L/4)^2}}\, ,
\label{smeq:Vcapprox}
\end{equation}
where $\epsilon_r$ is the dielectric constant of the material. The dimensionless Hamiltonian with the approximating formula used for the effective 1D Coulomb interaction reads as
\begin{equation}
 \frac{H^\text{rel}_2}{\Delta_o}= -\partial_x^2  +\frac 1 4 x^2 +  \frac{l_z}{a_B} \frac{1}{\sqrt{x^2+(L/4l_z)^2}}\, ,
\label{smeq:Hreldimless}
\end{equation}
where $x = z/l_z$. The Hamiltonian depends on two dimensionless parameters $l_z/a_B$ and $L/l_z$ (or equivalently $l_z/a_B$ and $L/a_B$). Therefore all the matrix elements in Eq.~\eqref{smeq:ZFS} can be expressed as a function of these quantities leading to $\zfsd =\eta\, m^* v_3^2 \hbar^4/l_z^4  $. The dimensionless coefficient $\eta$ depends on the relative length scales through the eigenstates of $H^\text{rel}_2$ and can be written as
\begin{eqnarray}
\begin{split}
\eta =& \frac{11}{16}\! -\! \frac{9}{2} \braket{\partial_x^2}_{_{o_1o_1}} \\
&+\!  \sum_i \!  \left \{  9\frac{|\braket{\partial_x^2}_{o_1o_i}|^2}{1+ \tilde \varepsilon_{o_i}\!  -\!  \tilde \varepsilon_{o_1} }\!  -\! 6 \frac{\text{Re}[ \braket{\partial_x^3}_{o_1 e_i}\! \braket{\partial_x}_{e_i o_1} ]}{\tilde \varepsilon_{e_i} - \tilde \varepsilon_{o_1}} \! - \! 4 \frac{|\! \braket{\partial_x^3}_{o_1 e_i}\!|^2 }{\tilde \varepsilon_{e_i} \! - \! \tilde \varepsilon_{o_1}}  \! -\! \frac{9}{4}   \frac{|\! \braket{\partial_x}_{o_1 e_i}\!|^2}{\tilde \varepsilon_{e_i} - \tilde \varepsilon_{o_1}}\!  -\! \frac{9}{2}  \frac{ |\! \braket{\partial_x}_{o_1 e_i}\!|^2}{2 +\tilde \varepsilon_{e_i} - \tilde \varepsilon_{o_1}}\!  \right \},
\end{split}
\end{eqnarray}
where $H^\text{rel}_2/\Delta_o \ket{e_i} = \tilde \varepsilon_{e_i} \ket{e_i}$ for the even, $ H^\text{rel}_2/\Delta_o \ket{o_i} = \tilde \varepsilon_{o_i} \ket{o_i}$ for the odd eigenstates with respect to $x$, and $-i\braket{\partial_x}_{nm} = -i\braket{\psi_n|\partial_x|\psi_m}$ is the matrix element of the dimensionless momentum. The coefficient $\eta$ is shown as a function of the two relative length scales $l_z/a_B$ and $L/a_B$ in Fig.~\ref{smfig:ZFScoeff_function}(a). Importantly, the dependence on the 1D cutoff $L/4$ is rather weak for $l_z \gtrsim a_B$ and therefore we expect the analytical formula provided in the main text, see Eq.~\eqref{eq:ZFSD}, to be valid for a wide range of cross sections.

The (unperturbed) exchange splitting $\ex_0 =\Delta_o ( \tilde \varepsilon_{o_1} - \tilde \varepsilon_{e_1})$ can also be expressed in terms of the two relative length scales $l_z/a_B$ and $L/a_B$, as $\ex_0 = \zeta \hbar^2 a_B^2 / m^* l_z^4$, where $\zeta$ is a dimensionless coefficient. Combining $\ex_0$ with $\zfsd$, we find that the anisotropy can be expressed as
\begin{equation}
\frac \zfsd \ex = \frac{m^{*2} v_3^2 \hbar^2}{a_B^2} \frac \eta \zeta+\mathcal O (v^4_3)\, .
\end{equation}
This equation shows that the anisotropy depends strongly on both $v_3$ and on the effective mass, i.e., $\zfsd/\ex \sim (m^*)^4$. The mass dependence can be understood by considering that $a_B \propto 1/m^*$ and that $\eta/\zeta$ depends weakly on $a_B$, and therefore on the mass as well [see Fig.~\ref{smfig:ZFScoeff_function}(b)].

\section{Details of the numerical calculation}
\label{sm:numcalc}

Here, we discuss in detail the numerical calculations introduced in the main text. We start the numerical analysis by considering a single QD with two holes, and assume harmonic confinement along the wire ($z$ direction) as
\begin{eqnarray}
H = H_{K,1}(\mathbf p_1)+H_{K,2}(\mathbf p_2) + \frac{\hbar^2 \gamma_1}{2 m_e L^4_z}(z_1^2+z_2^2) +C(\mathbf r_1  - \mathbf r_2)\, ,
\label{smeq:H2partfull}
\end{eqnarray}
where $\mathbf p_i$ and $\mathbf r_i$ are the momentum and spatial coordinate of the $i$th particle, $C(\mathbf r_1  - \mathbf r_2)$ is the Coulomb interaction, and hard-wall boundary conditions in the $x$-$y$ directions are implied. The $n\times n$ Kane model describing $n =4$ or $n=6$ valence bands in inversion symmetric semiconductors close to the $\Gamma$ point is~\cite{winkler:book}
\begin{eqnarray}
\begin{split}
H^{n\times n}_{K} (\bf p)=& \sum_{\alpha = 1 }^n E_\alpha \ket \alpha \bra \alpha +\frac{\gamma_1}{2m_e}  p^2 -\left( \frac{\gamma_2}{m_e}  p^2_x  + \frac 2 3  D_u \epsilon_{xx} \right) A_{xx}  - \left( 2\frac{\gamma_3}{m_e} \{p_x , p_y\}  + \frac 4 3 D'_u \epsilon_{xy} \right)A_{xy}+ c.p.  \, ,
 \end{split}
 \label{smeq: nxnKane}
\end{eqnarray}
where $E_\alpha$ is the energy of the band $\alpha$ at ${\bf p}=0$, the coefficients $\gamma_1$, $\gamma_2$, and $\gamma_3$ are the Luttinger parameters determined by the band structure of the material, $D_u$ and $D'_u$ are the deformational potentials, and $A_{ij}$ are $n\times n$ matrices acting on the band degree of freedom. Moreover, the anticommutator between two operatrs $O_1$ and $O_2$ is defined as $\{O_1,O_2\} = (O_1O_2+O_2O_1)/2$. For example in the $4$-band Luttinger-Kohn model describing the top of the HH and LH bands, $E_\alpha = 0$ and $A_{ij} = \{J_i,J_j\}$ where $J_i$ are the spin-$3/2$ matrices. For the $6$-band model --obtained by considering the third pair of valence bands-- the splitt-off holes are shifted by $\Delta_{SO}$ from the HH and LH bands, and the $A_{ij}$ matrices are given in Ref.~\cite{winkler:book}. Throughout this work, we assume compressive strain with strain tensor $\epsilon_{ij} \propto \epsilon_{ii} \delta_{ij}$.

\subsection{Long-QD calculation}
\label{sm:longQD}

Here we study the long QD case, where the Coulomb interaction is weak compared to the transversal confinement energy but stronger than the longitudinal confinement energy, i.e., $ L/\pi \lesssim a_B \lesssim L_z$. We start by deriving an effective 1D model (in $z$ direction) accounting for a few NW subbands. To this goal, we add the electric field term $H_E=-e{\bf E \cdot r}$ to the Hamiltonian of Eq.~\eqref{smeq: nxnKane}, impose the hard-wall boundary conditions in the $x-y$ directions, and expand the full Hamiltonian in powers of momentum $p_z$. The operator multiplying $p_z^j$ reads as $H^{(j)}_K = \frac 1 {j!} \partial^j_{p_z} H_{K} ({\bf p})\big |_{p_z=0}$ for $j \in \{0,1,2\}$~\cite{bosco2021sm}. Then, we find the eigensystem of the Hamiltonian $H^{(0)}_{K}$ as
\begin{equation}
H^{(0)}_{K}\phi_n (x,y,s) = \varepsilon_n \phi_n (x,y,s)\, ,
\end{equation}
where $s$ is the band index of the Kane model. The eigenstates $\phi_n (x,y,s)$ include the effects of electric field and strain and are used to project the full Hamiltonian onto the 1D subspace as
\begin{equation}
[H_{K} (p_z)]_{nm} = \varepsilon_n \delta_{nm} +  [H^{(1)}_{K}]_{nm} p_z +  [H^{(2)}_{K}]_{nm} p_z^2\, ,
\label{smeq:HKnoSWT}
\end{equation}
where the indices $m$ and $n$ label the NW subbands. We include a large number of NW subbands ($N_{xy} = 200$ in the present work) and we derive the effective wire model
\begin{equation}
[\tilde H_{K} (p_z)]_{nm} = \varepsilon_n \delta_{nm} +  [H^{(1)}_{K}]_{nm} p_z +  [\tilde H^{(2)}_{K}]_{nm} p_z^2 +  [\tilde H^{(3)}_{K}]_{nm} p_z^3 + \mathcal O (p_z^4)\, ,
\label{smeq:HKSWT}
\end{equation}
by third order SW transformation.  By using this effective Hamiltonian instead of Eq.~\eqref{smeq:HKnoSWT}, we can restrict ourselves to a few number of bands, greatly simplifying the two-body problem.

In our numerical analysis we applied an additional transformation that helps to improve the convergence of the ZFSs for large linear SOI. For this, we divide the Hamiltonian in Eq~\eqref{smeq:HKSWT} into $2\times 2$ blocks according to the Kramers partners. Due to time reversal symmetry, each diagonal block has to be of the form of Eq.~\eqref{smeq:HNW}, therefore for each subband one can apply a spin dependent momentum shift analogous to the one in Eq.~\eqref{smeq:HNWSOI}.

Using the NW subbands, the two particle Hamiltonian of Eq.~\eqref{smeq:H2partfull} reads
\begin{eqnarray}
H_{m_1,m_2}^{n_1,n_2} (z_1,z_2) = [H_{K} (p_{z_1})]_{n_1m_1} +[H_{K} (p_{z_2})]_{n_2m_2}  + \frac{\hbar^2 \gamma_1}{2 m_e L^4_z}(z_1^2+z_2^2) +C_{m_1,m_2}^{n_1,n_2} (z_1-z_2)\, ,
\label{smeq:H2strong}
\end{eqnarray}
where the Coulomb matrix elements are defined as $C_{m_1,m_2}^{n_1,n_2}  = \braket{\phi_{n_1},\phi_{n_2}|C|\phi_{m_1},\phi_{m_2}}$. For example the Coulomb matrix element in the lowest subbands is $C_{m_1,m_2}^{n_1,n_2} (z_1-z_2)  = \delta_{m_1,n_1}  \delta_{m_2,n_2} V_c (z_1-z_2)$, (where $m_{1,2},n_{1,2} \in \{1,2\}$) that has no singularity at $z_1 = z_2$, and is well approximated by using a simple cutoff determined by the transversal confinement length as shown in Eq.~\eqref{smeq:Vcapprox}. 

To diagonalize Eq.~\eqref{smeq:H2strong}, we move to the COM frame, using the relation in Eq.~\eqref{smeq:COMcoord}, and we define the orthonormal basis states 
\begin{equation}
\psi^{n_1,n_2}_{u,w, s_1,s_2} ({\bf r}_1, {\bf r}_2) = \phi_{n_1} (x_1,y_1,s_1)\, \phi_{n_2} (x_2,y_2,s_2)\, \phi^\text{\tiny{COM}}_{n_1,n_2,u} \left[\frac 1 2 (z_1+z_2)\right]\, \phi^\text{rel} _{n_1,n_2,w}(z_1-z_2)\, .
\end{equation}

The COM basis state $\phi^\text{\tiny{COM}}$ satisfy the  eigenvalue equation $H^\text{\tiny{COM}} (n_1,n_2) \phi^\text{\tiny{COM}}_{n_1,n_2,u} (Z) = \varepsilon^\text{\tiny{COM}}_u \phi^\text{\tiny{COM}}_{n_1,n_2,u} (Z)$, with the Hamiltonian
\begin{equation}
H^\text{\tiny{COM}} (n_1,n_2) = \frac 1 4 \left( [\tilde H^{(2)}_{K}]_{n_1n_1}  + [\tilde H^{(2)}_{K}]_{n_2n_2} \right) k_Z^2  + \frac{\hbar^2 \gamma_1}{m_e L_z^4} Z^2\, .
\end{equation}
We note that the COM basis states are harmonic oscillator eigenstates with subband dependent mass $1/m_{n_1, n_2}= \left( [\tilde H^{(2)}_{K}]_{n_1n_1}  + [\tilde H^{(2)}_{K}]_{n_2n_2} \right) / 2 \hbar^2$. In contrast, the basis states of the relative coordinate depend also on the Coulomb potential. We use $\phi^\text{rel}$ as basis states, i.e., the eigenfunctions satisfying the eigenvalue equation $H^\text{rel}  (n_1,n_2) \phi^\text{rel} _{n_1,n_2,w}(z) = \varepsilon^\text{rel} _w \phi^\text{rel} _{n_1,n_2,w}(z)$, with Hamiltonian
\begin{equation}
H^\text{rel} (n_1,n_2)= \left( [\tilde H^{(2)}_{K}]_{n_1n_1}  + [\tilde H^{(2)}_{K}]_{n_2n_2} \right) k_z^2  + \frac{\hbar^2 \gamma_1}{4m_e L_z^4} z^2 + C_{n_1,n_2}^{n_1,n_2} (z)\, .
\end{equation}

\subsection{Short-QD calculation}
\label{sm:shortQD}

Using a few NW subbands as a basis for the numerical calculation is only justified if the longitudinal confinement length is large compared to the width of the NW, i.e., $l_z > L/\pi$. However, in short QDs, where the ZFS is expected to be stronger, several subbands may be required. In this case, instead of effective wire bands, we use multiple basis states in the $x$-$y$ direction that satisfy the appropriate boundary conditions. In the present work, we start from the $4 \times 4$ Kane model and use the particle in a box basis states
\begin{equation}
\phi_{n, m} (x,y) \ket{3/2, s} = \frac 2 L \cos (n \pi x /L) \cos (m \pi y /L) \ket{3/2, s}\, ,
\end{equation}
where the spin part is $\ket{3/2, s} \equiv \ket{j=3/2,j_z=s}$. Along the $z$ direction, the  COM and relative coordinate basis states are chosen as eigenstates of the Hamiltonians
\begin{equation}
H^\text{\tiny{COM}}_{s_1,s_2}= \frac{\hbar^2}{2m_s} k_Z^2  + \frac{\hbar^2 \gamma_1}{m_e L_z^4} Z^2\, ,
\end{equation}
\begin{equation}
H^\text{rel}_{s_1,s_2} = \frac{2\hbar^2}{m_s} k_z^2  + \frac{\hbar^2 \gamma_1}{4m_e L_z^4} z^2 + C_{n_1,m_1,n_2,m_2}^{n_1,m_1,n_2,m_2} (z)\, ,
\end{equation}
respectively, where the mass is 
\begin{equation}
\frac 1 {m_s} = \frac 1 {m_e} \begin{cases}
(\gamma_1 - 2 \gamma_2)\, , &\text{ if } |s_1| = |s_2| = 3/2\\
(\gamma_1 + 2 \gamma_2)\, , &\text{ if } |s_1| = |s_2| = 1/2\\
\gamma_1\, , &\text{ if } |s_1| \neq |s_2|
 \end{cases}\, .
\end{equation}
Finally, the resulting the two-particle basis states used to diagonalize the complete 2-body Hamiltonian are
\begin{equation}
\psi^{n_1,m_1,n_2,m_2}_{u,w,s_2,s_2} ({\bf r}_1, {\bf r}_2) =\phi_{n_1, m_1} (x_1,y_1) \phi_{n_2, m_2} (x_2,y_2)  \phi^\text{\tiny{COM}}_{s_1,s_2,u} \left[\frac 1 2 (z_1+z_2)\right]\, [\phi^\text{rel}]^{n_1,m_1,n_2,m_2}_{s_1,s_2,w}(z_1-z_2)\, .
\end{equation}

\subsection{Anisotropic short-range corrections to the Coulomb interaction}

In Ref.~\cite{secchi:arx20} it is shown that the Coulomb interaction can acquire anisotropic corrections at short distances that couple the band degrees of freedom, i.e., the HH, LH, and the spin-orbit split-off bands. This effect is a consequence of the finite orbital angular momentum of the $p$-type wavefunctions corresponding to the valence bands.

Three different type of corrections were identified in Ref.~\cite{secchi:arx20}: intraband, partially intraband, and interband corrections. The intraband and partially intraband terms contain both short-range ($r<a/4$, where $a$ is the lattice constant) and long-range ($a/4 < r \lesssim 2a$) contributions, while the interband corrections are exclusively short-ranged. Here we omit the long-ranged contributions as their contribution is negligible compared to the short-range terms~\cite{secchi:arx20}. The form of the short-range Coulomb corrections used in our work is
\begin{eqnarray}
\begin{split}
\delta C_\text{s-r} =& \frac{F_2}{25}\, g_d(\mathbf r_1- \mathbf r_2) \left[ P_\text{\tiny{HH}}(1)P_\text{\tiny{HH}}(2)+P_\text{\tiny{LH}}(1) P_\text{\tiny{LH}}(2)-P_\text{\tiny{LH}}(1)P_\text{\tiny{HH}}(2)-P_\text{\tiny{HH}}(1)P_\text{\tiny{LH}}(2) \right]\\
&+\sqrt 2 \frac{F_2}{25}\, g_d(\mathbf r_1- \mathbf r_2) [J_\text{part,\,d} (1) J_\text{part,\,od} (2) + J_\text{part,\,od} (1) J_\text{part,\,d} (2)]\\
&+\frac{F_2}{25}\, g_d(\mathbf r_1- \mathbf r_2) \left[ J_\text{part,\,od} (1) J_\text{part,\,od} (2) + 3 J_\text{int,\,Y}^{} J_\text{int,\,Y}^\dagger + 3 J_\text{int,\,Y}^\dagger  J_\text{int,\,Y}^{}  + 6 J_\text{int,\,X}^{} J_\text{int,\,X}^\dagger + 6 J_\text{int,\,X}^\dagger  J_\text{int,\,X}^{} \right] ,
\end{split}
\label{smeq:Cbshort}
\end{eqnarray}
where the first term is  the intraband, the second term is the partially interband, the third term is the interband correction, and $F_2 = F_2(4p,4p) = 4.235\,$eV is the relevant Slater-Condon parameter for Ge as provided in Ref.~\cite{satoko:sc90}.
The functional form of $g_d(\mathbf r)$ has been derived for the continuum representation of the atomistic model in Ref.~\cite{secchi:arx20}. Here we provide only the simplest approximation for this short-ranged function, i.e.,
\begin{equation}
g_d(\mathbf r) \propto \left(\frac{a}{2}\right)^3 \delta(\mathbf r)\, .
\end{equation} 
Since $g_d(\mathbf r) $ is cut at the boundary of a cube with an edge of $a/2$ (where $a =0.56\,$nm is the lattice constant for Ge), the spatial dependence is well approximated by a Dirac delta within the envelope function approximation (i.e., $L,l_z\gg a$).

In order to simplify the formulas of Ref.~\cite{secchi:arx20} to the case of the $6\times 6$ Kane model in Eq.~\eqref{smeq:Cbshort}, we introduced the following operators
\begin{subequations}
\begin{equation}
P_\text{\tiny{HH}} = \ket{\frac 3 2 ,\, \frac 3 2}\bra{\frac 3 2 ,\, \frac 3 2} 
+ \ket{\frac 3 2 ,\, -\frac 3 2}\bra{\frac 3 2 ,\, -\frac 3 2}\, ,
\end{equation}
\begin{equation}
P_\text{\tiny{LH}} = \ket{\frac 3 2 ,\, \frac 1 2}\bra{\frac 3 2 ,\, \frac 1 2} 
+ \ket{\frac 3 2 ,\, -\frac 1 2}\bra{\frac 3 2 ,\, -\frac 1 2}\, ,
\end{equation}
\begin{equation}
J_\text{part,\,d} = P_\text{\tiny{HH}}-P_\text{\tiny{LH}}\, ,
\end{equation}
\begin{equation}
J_\text{part,\,od} =  \ket{\frac 3 2 ,\, \frac 1 2}\bra{\frac 1 2 ,\, \frac 1 2} 
+ \ket{\frac 3 2 ,\, -\frac 1 2}\bra{\frac 1 2 ,\, -\frac 1 2} 
+ h.c.\, ,
\end{equation}
\begin{equation}
J_\text{int,\,X} = -\frac 1 {\sqrt 3}\ket{\frac 3 2 ,\, \frac 3 2}\bra{\frac 3 2 ,\, -\frac 1 2} 
-\sqrt{\frac 2  3}\ket{\frac 3 2 ,\, \frac 3 2}\bra{\frac 1 2 ,\, -\frac 1 2} 
+ \frac 1 {\sqrt 3}\ket{\frac 3 2 ,\, \frac 1 2}\bra{\frac 3 2 ,\, -\frac 3 2} 
+ \sqrt{\frac 2  3}\ket{\frac 1 2 ,\, \frac 1 2}\bra{\frac 3 2 ,\, -\frac 3 2}\, ,
\end{equation}
\begin{eqnarray}
\begin{split}
J_\text{int,\,Y} =-\sqrt{\frac 2  3} \ket{\frac 3 2 ,\, \frac 3 2}\bra{\frac 3 2 ,\, \frac 1 2} 
+ \frac 1 {\sqrt 3}\ket{\frac 3 2 ,\, \frac 3 2}\bra{\frac 1 2 , \frac 1 2} 
-\sqrt{\frac 2  3}\ket{\frac 3 2 ,\, -\frac 1 2}\bra{\frac 3 2 ,-\frac 3 2} 
+ \frac 1 {\sqrt 3}\ket{\frac 1 2 ,-\frac 1 2}\bra{\frac 3 2 , -\frac 3 2} \\
+\ket{\frac 3 2 ,\, \frac 1 2}\bra{\frac 1 2 , -\frac 1 2}
-\ket{\frac 1 2 ,\, \frac 1 2}\bra{\frac 3 2 , -\frac 1 2}  \, ,
\end{split}
\end{eqnarray}
\end{subequations}
where the states $\ket{j,j_z}$ are eigenstates of the total angular momentum operators $\hbar^2 \hat{J}^2$ and $\hbar \hat{J}_z$ with eigenvalues $\hbar^2 j(j+1)$ and $\hbar j_z$, respectively.

\subsection{Comparison between short and long QDs}

\begin{figure}
\includegraphics[width=0.8\textwidth]{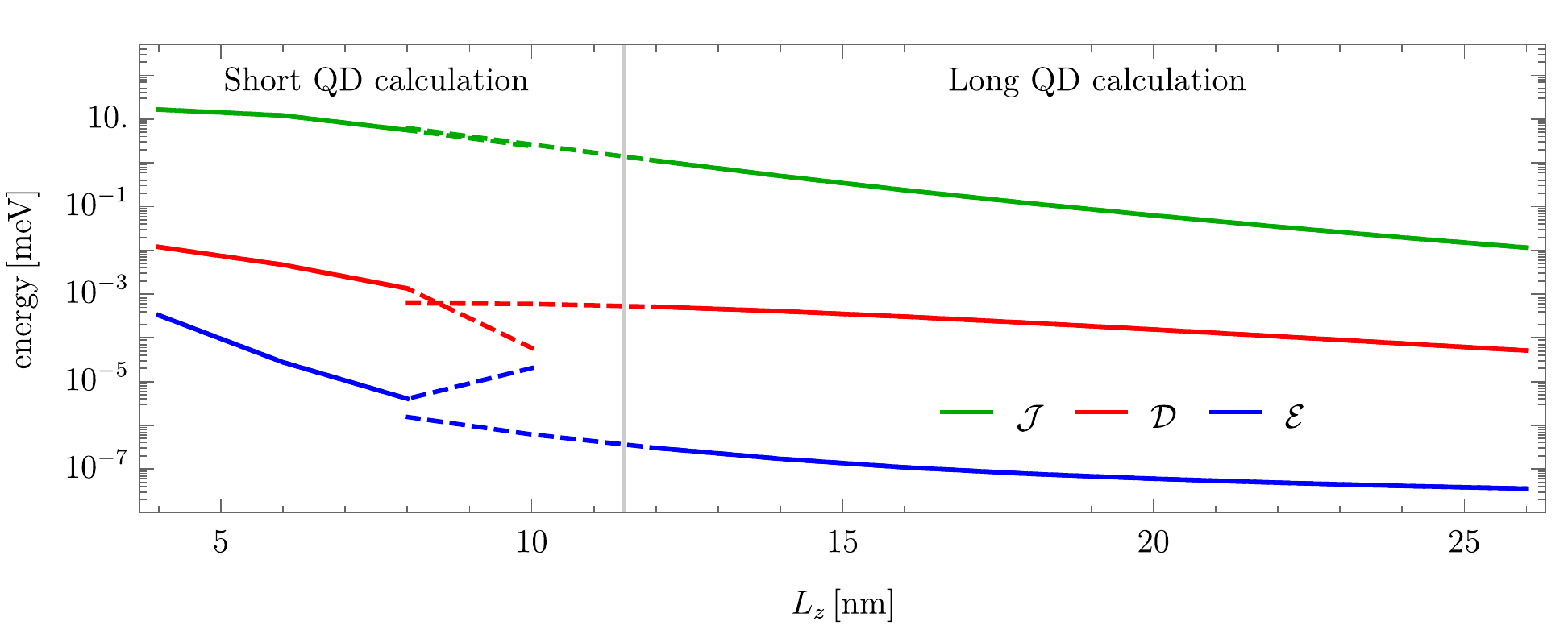}
\caption{Exchange splitting $\ex$ and zero-field splittings $\zfsd$ and $\zfse$ as a function of $L_z$ in a Ge quantum dot. We consider a square wire with side length $L= 10\,$nm, compressive strain $\epsilon_{zz}= -2.5\%$, and electric field $E_x = 2\,$V$/\mu$m. The $\{x,y,z\}$ axes of the wire correspond to the $\braket{100}$ crystallographic directions. The first set of curves starting from $L_z = 4\,$nm to $L_z = 10\,$nm are calculated in the short QD assumption discussed in Sec.~\ref{sm:shortQD}, while the second set from $L_z = 8\,$nm to $L_z = 26\,$nm is calculated using the long QD calculation discussed in Sec.~\ref{sm:longQD} and also used in the main text. The vertical line corresponds to $L_z = a_B$ in Ge.} 
\label{smfig:weakvsstrong}
\end{figure}

In this section we compare the two numerical approaches described in Secs.~\ref{sm:longQD} ~and~\ref{sm:shortQD}  to calculate the exchange- and zero-field splittings. The first approach well describes long quantum dots, with $L_z>a_B, L$. In this approach we account for 4 NW subbands, and 30 states for the COM and 30 states for the relative coordinates. The second approach works for short QDs, with $L_z~L<a_B$, and uses basis states adapted to the confinement in each spatial directions ($3\,$-$\,3$ particle in a box eigenstates in $x-y$ and $8$ harmonic oscillator eigenstates in $z$ directions) and therefore describe the short QD limit, i.e., $L_z \sim L < a_B$. Since the numerical analysis in short QDs requires a large number of basis states to converge, in these calculations we omit the spin-orbit split-off bands (reducing the size of the two-particle Hilbert space to $82'944$ in the short QD case). The effect of the split-off holes is fully accounted for in the main text. 

The results of the two numerical solutions are compared in Fig.~\ref{smfig:weakvsstrong} for a Ge wire with square cross-section with side length $L= 10\,$nm, compressive strain $\epsilon_{zz}= -2.5\%$, and electric field $E_x = 2\,$V$/\mu$m. The $\{x,y,z\}$ axes of the wire correspond to the $\braket{100}$ crystallographic directions. The simulation of the exchange shows a good quantitative agreement in the two cases. The ZFSs computed in these cases are also in qualitative agreement, however, at $L \sim 10\,$nm the numerical precision used for short QDs is not sufficient and the results of the short QD simulation are not reliable for larger QD lengths. We expect that the ZFS interpolates smoothly between the two limits.

From this comparison we conclude that  the results obtained with the long quantum dot procedure remain reasonably accurate even at rather small values of $L_z$, confirming also the numerical and analytical theory discussed in the main text.
 
\section{Symmetry analysis of the triplet degeneracy}
\label{sm:tripletdeg}

\begin{table}
\begin{tabular}{c|c c c c c }
$O_h$ & \hspace{0.3cm} $\Gamma^+_1(1)$\hspace{0.3cm} & \hspace{0.3cm}$\Gamma^+_2(1)$\hspace{0.3cm} & \hspace{0.3cm}$\Gamma^+_3(2)$\hspace{0.3cm} & \hspace{1.4cm}$\Gamma^+_4(3)$\hspace{1.4cm} & \hspace{1.3cm}$\Gamma^+_5(3)$\hspace{1.3cm} \\ 
\hline 
$D_{4h}$ & $\Gamma^+_1(1)$ & $\Gamma^+_3(1)$ & $\Gamma^+_1(1)+\Gamma^+_3(1)$ & $\Gamma^+_2(1)+\Gamma^+_5(2)$ & $\Gamma^+_4(1)+\Gamma^+_5(2)$ \\ 
$D_{2h}$ & $\Gamma^+_1(1)$ & $\Gamma^+_1(1)$ & $2\Gamma^+_1(1)$ & $\Gamma^+_2(1)+\Gamma^+_3(1)+\Gamma^+_4(1)$ & $\Gamma^+_2(1)+\Gamma^+_3(1)+\Gamma^+_4(1)$ \\ 
$C_{2v}$ & $\Gamma_1(1)$ & $\Gamma_2(1)$ & $\Gamma_1(1)+\Gamma_2(1)$ & $\Gamma_2(1)+\Gamma_3(1)+\Gamma_4(1)$ & $\Gamma_1(1)+\Gamma_3(1)+\Gamma_4(1)$ \\  
\end{tabular}
\caption{Compatibility table of the cubic point group~\cite{koster:book}. For NW QDs the symmetry groups $D_{4h}$, $D_{2h}$, and $C_{2v}$ correspond to square, rectangular cross section, and the hut wire, respectively. Assuming that the confinement along the wire is symmetric, the coordinate axes $x,y,z$ correspond to $\braket{100}$ crystallographic axes, and no additional fields are applied.}
\label{tab:Ohcomp} 
\end{table} 

In this section we use group theoretical tools to study the degree of degeneracy of the two-particle eigenstates that is allowed by the irreducible representations of the two-particle point groups (i.e., double groups). Starting from the case with cubic symmetry, we consider the compatibility table of the cubic point group and we show how the degeneracy is resolved if certain symmetries are broken by e.g., the interface, electric field, or strain.

In the following discussion, we restrict our attention to the HH and LH bands. These bands at ${\bf k} = 0$ are described by the irreducible representation $\Gamma^+_8(4)$, where $"+"$ indicates even parity with respect to inversion and the number in parentheses is the dimension of the representation i.e., the degree of degeneracy~\cite{solyom2007sm}. By assuming the most general form of the interaction --e.g., accounting for the short-range interband Coulomb interaction-- the two-particle representation can be decomposed into irreducible representations as follows
\begin{equation}
\Gamma^+_8 (4) \times \Gamma^+_8 (4) = \Gamma^+_1(1) +\Gamma^+_2(1) +\Gamma^+_3(2) + 2\Gamma^+_4(3)+ 2\Gamma^+_5(3)\, ,
\end{equation}
where one obtains 1-, 2-, and 3-dimensional irreducible representations. This decomposition implies that the full 3-fold degeneracy of the triplet states is maintained if the QD confinement respects every symmetry of the cubic point group. In experiments this is usually not the case, therefore we consider the few nontrivial point groups that are of practical relevance:
\begin{itemize}
\item[(i)] a NW with square cross section and $E_x = 0$ as in Figs.~\ref{fig:ZFS_Lz_Ex}~and~\ref{fig:ZFS_Ge_Si} of the main text, described by the $D_{4h}$ tetragonal point group that contains one fourfold and two twofold rotation axes as well as inversion symmetry. The triplet degeneracy is indeed lifted as predicted by the first line of Tab.~\ref{tab:Ohcomp}.
\item[(ii)] a rectangular NW in the absence of electric field, or a square wire with compressive strain along the $x$ or $y$ directions, described by the $D_{2h}$ orthorhombic point group that contains three twofold rotation axes as well as inversion symmetry. In this case each of the three triplets are non-degenerate (see second line of Tab.~\ref{tab:Ohcomp}). This case has been confirmed in our numerical calculation (not shown).
\item[(iii)] a rectangular or square NW with electric field applied perpendicular to either of the sides of the cross section, or a NW with an equilateral triangle cross section. These cases are both  described by the $C_{2v}$ orthorhombic point group that contains two reflection planes and one twofold rotation axis. In this case each of the three triplets are non-degenerate (see third line of Tab.~\ref{tab:Ohcomp}). This case has been confirmed as well by our numerical calculation (see Figs.~\ref{fig:ZFS_Lz_Ex}~and~\ref{fig:ZFS_Ge_Si} of the main text).
\end{itemize}

Finally, we note that including only the spin-independent  Coulomb interaction is not enough to lift all the triplet degeneracies as predicted by symmetries. To obtain the lowest possible degeneracy, short-range interband corrections to the Coulomb interaction also need to be considered. However, in the main text, we show that these effects are significantly smaller than the cubic spin-orbit induced lifting of triplet degeneracy.

\bibliographystyle{unsrt}

\end{document}